\documentclass[prd,twocolumn,tightenlines,superscriptaddress,nofootinbib,
showpacs]{revtex4}
\usepackage{amssymb,latexsym}
\usepackage{amsmath,amsbsy,bbm}
\usepackage{epsfig,bm}
\usepackage{graphicx,comment}
\usepackage{color}
\begin{document} 

\preprint{MIT-CTP-4080}
\hfill
MIT-CTP-4080

\title{The Phase Transition of the 
Spin-1/2 Heisenberg Model with a Spatially Staggered Anisotropy on the
Square Lattice.}

\author{F.-J. Jiang}
\email[]{fjjiang@mit.edu}
\affiliation{Center for Theoretical Physics, Massachusetts Institute of Technology,
77 Massachusetts Ave, Cambridge, MA 02139}

\vspace{-1cm}
  
\begin{abstract}
Puzzled by the indication of a new critical theory for the
spin-1/2 Heisenberg model with a spatially staggered anisotropy on the
square lattice as suggested in \cite{Wenzel08}, we re-investigate
the phase transition of this model induced by dimerization.
We focus on studying the finite-size scaling of the observables 
$\rho_{s1} L$ and $\rho_{s2} L$, where $L$ stands for the spatial box sizes used in the 
simulations and $\rho_{si}$ with $i \in \{1,2\}$ is the 
spin-stiffness in $i$-direction. We find by performing finite-size scaling using the observable $\rho_{s2} L$,
which corresponds to the spatial direction with a fixed antiferromagnetic 
coupling, one would suffer a much less severe correction compared to that of using
$\rho_{s1} L$. Therefore $\rho_{s2} L$ is a better quantity than $\rho_{s1} L$ for
finite-size scaling analysis concerning the limitation for the availability of 
large volumes data in our study. Remarkably, by employing the method of fixing the 
aspect-ratio of spatial winding numbers squared in the simulations, 
even from $\rho_{s1} L$ which receives 
the most serious correction among the observables
considered in this study, we arrive at a value for the critical exponent 
$\nu$ which is consistent with the expected $O(3)$ value by using only up
to $L = 64$ data points.
\end{abstract}

\maketitle

\section{Introduction}
\vskip-0.2cm
Heisenberg-type models have been studied in great detail
during the last twenty years because of their 
phenomenological importance. 
For example, it is believed that the spin-1/2 Heisenberg model on the square 
lattice is the correct model for understanding the undoped precursors of 
high $T_c$ cuprates (undoped antiferromagnets).
Further, due to the availability of efficient Monte Carlo algorithms as well
as the increasing power of computing resources, properties of undoped antiferromagnets
on geometrically non-frustrated lattices have been determined to unprecedented 
accuracy \cite{Sandvik97,Sandvik99,Kim00,Wang05,Jiang08,Alb08,Wenzel09}. 
For instance, using a loop algorithm, the low-energy
parameters of the spin-1/2 Heisenberg model on the square lattice
are calculated very precisely and are in quantitative agreement with the experimental 
results \cite{Wie94}. Despite being well studied, 
several recent numerical investigation of anisotropic Heisenberg models have led to unexpected
results \cite{Wenzel08,Pardini08,Jiang09.1}. In particular, Monte Carlo evidence indicates that the anisotropic
Heisenberg model with staggered arrangement of the antiferromagnetic 
couplings may belong to a new universality class, in contradiction
to the theoretical $O(3)$ universality prediction \cite{Wenzel08}.
For example, while the most accurate Monte Carlo value for the critical exponent
$\nu$ in the $O(3)$ universality class is given by $\nu=0.7112(5)$ \cite{Cam02},
the corresponding $\nu$ determined in \cite{Wenzel08} is shown to be $\nu=0.689(5)$. 
Although subtlety of calculating the critical exponent $\nu$ from performing
finite-size scaling analysis is demonstrated for a similar anisotropic 
Heisenberg model on the honeycomb lattice \cite{Jiang09.2}, the discrepancy between $\nu = 0.689(5)$ 
and $\nu=0.7112(5)$ observed in \cite{Wenzel08} remains to be understood.

In order to clarify this issue further, we have simulated the spin-1/2 Heisenberg model with
a spatially staggered anisotropy on the square lattice. Further, we choose to analyze 
the finite-size scaling of the observables $\rho_{s1} L$ and $\rho_{s2} L$,
where $L$ refers to the box sizes used in the simulations and $\rho_{si}$ with 
$i \in \{1,2\}$ is the spin stiffness in $i$-direction.
The reason for choosing $\rho_{s1} L$ and $\rho_{s2} L$ is twofold. 
First of all, these two observables can be calculated
to a very high accuracy using loop algorithms. Secondly, one can measure 
$\rho_{s1}$ and $\rho_{s2}$ separately. In practice, one would
naturally either measure $\rho_s$ which is the average of $\rho_{s1}$ and $\rho_{s2}$ 
in order to increase the statistics, or $\rho_{s2}$, which
corresponds to the spatial direction with a fixed antiferromagnetic coupling, 
would not be used for data analysis since more measurements is required in order 
to obtain a good statistics 
for this observable. However for the model considered here, 
it is useful to measure quantities which are sensitive to anisotropy.
Surprisingly, as we will show later, the observable $\rho_{s2} L$ receives a much
less severe correction than $\rho_{s1} L$ does.
Hence $\rho_{s2} L$ is a better observable than
$\rho_{s1} L$ (or $\rho_{s} L$) for finite-size scaling analysis concerning the limitation for
the availability of large volumes data in this study. 
In addition, instead of
using a fixed aspect-ratio of spatial box sizes as done in most Monte Carlo calculations, in our investigation 
we employ the method of fixing the
aspect-ratio of spatial winding numbers squared in the simulations which we will introduce 
briefly later.
Remarkably, combining the idea of fixing the aspect-ratio of spatial winding numbers squared in the 
simulations and finite-size scaling analysis, unlike the unconventional value for
$\nu$ observed in \cite{Wenzel08}, even from $\rho_{s1} L$ which suffers a very serious correction,
we arrive at a value for $\nu$ which is consistent with that of $O(3)$ by using only up to $L=64$ data points.  
   
This paper is organized as follows. In section \ref{model}, the anisotropic
Heisenberg model and the relevant observables studied in this work are briefly described.
Section \ref{results} contains our numerical results.
In particular, the corresponding critical point as well as the critical
exponent $\nu$ are determined by fitting the numerical data to their predicted
critical behavior near the transition. Finally, we conclude our study
in section \ref{discussion}.

\begin{figure}
\begin{center}
\includegraphics[width=0.33\textwidth]{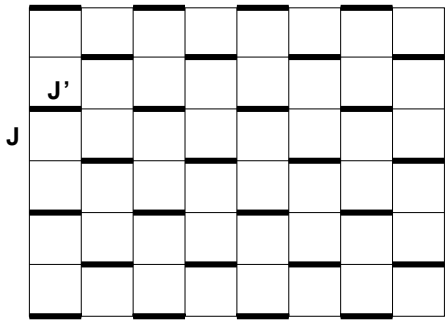}
\end{center}\vskip-0.5cm
\caption{The anisotropic Heisenberg model considered in this study.}
\label{fig2}
\end{figure}

\vskip-0.2cm

\section{Microscopic Model and Corresponding Observables}
\label{model}\vskip-0.2cm 
The Heisenberg
model considered in this study is defined by the Hamilton operator
\begin{eqnarray}
\label{hamilton}
H = \sum_{\langle xy \rangle}J\,\vec S_x \cdot \vec S_{y}
+\sum_{\langle x'y' \rangle}J'\,\vec S_{x'} \cdot \vec S_{y'},
\end{eqnarray}
where $J'$ and $J$ are antiferromagnetic exchange couplings connecting
nearest neighbor spins $\langle  xy \rangle$
and $\langle x'y' \rangle$, respectively. Figure 1 illustrates the Heisenberg
model described by Eq.~(\ref{hamilton}). 
To study the critical behavior of this anisotropic Heisenberg model near 
the transition driven by the anisotropy, in particular to determine 
the critical point as well as the critical exponent $\nu$, 
the spin stiffnesses in the $1$- and $2$-directions which are defined by\vskip-0.5cm
\begin{eqnarray}
\rho_{si} = \frac{1}{\beta L^2}\langle W^2_{i}\rangle,
\end{eqnarray}
are measured in our simulations.
Here $\beta$ is inverse temperature and $L$ 
refers to the spatial box sizes. Further $W^2_{i}$ is
the winding number squared in the $i$-direction.
By carefully investigating the spatial volumes 
and the $J'/J$ dependence of
$\rho_{s i}L$, one can determine the critical point as well
as the critical exponent $\nu$ with high precision.

\begin{figure}
\begin{center}
\vbox{
\includegraphics[width=0.375\textwidth]{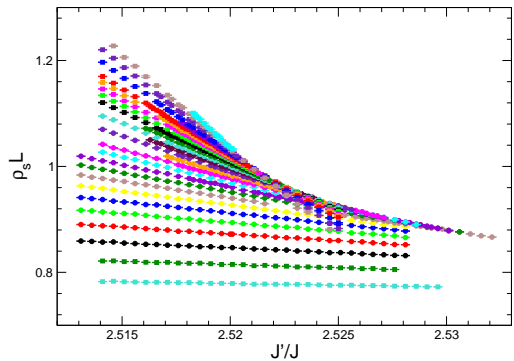}\vskip1.0cm
\includegraphics[width=0.375\textwidth]{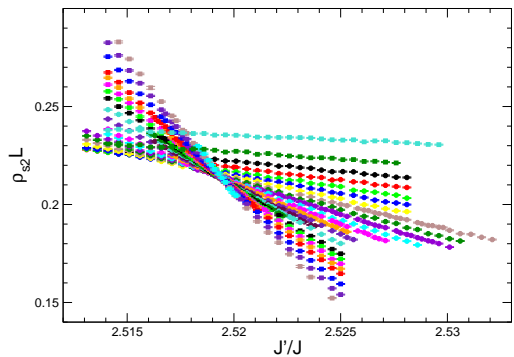}
}
\end{center}\vskip-0.5cm
\caption{$\rho_s L$ (upper panel) and $\rho_{s2} L$ (lower panel) as 
functions of $J'/J$.}
\label{fig0}
\end{figure}

\begin{figure}
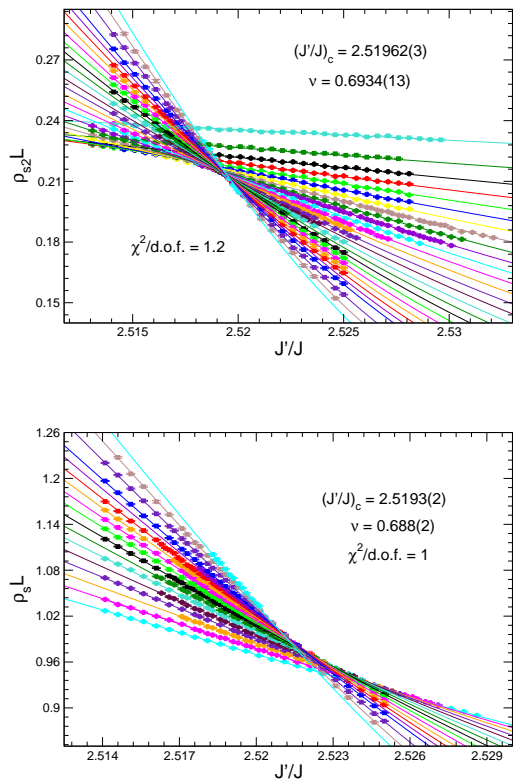

\begin{center}
\vbox{
\includegraphics[width=0.375\textwidth]{rhos2L2fit.1.eps}\vskip0.9cm
\includegraphics[width=0.375\textwidth]{rhosLfit.1.eps}\vskip0.9cm
}
\end{center}\vskip-0.5cm
\caption{Fits of $\rho_{s} L$ (upper panel) and $\rho_{s2}L $ 
(lower panel) to Eq.~\ref{FSS}. While the circles and squares on these two panels
are the numerical Monte Carlo data from the simulations, the solid curves 
are obtained by using the results from the fits.}
\label{fig1}
\end{figure}

\vskip-0.2cm

\section{Determination of the Critical Point and the Critical Exponent $\nu$}
\label{results}
\begin{figure}
\begin{center}
\includegraphics[width=0.365\textwidth]{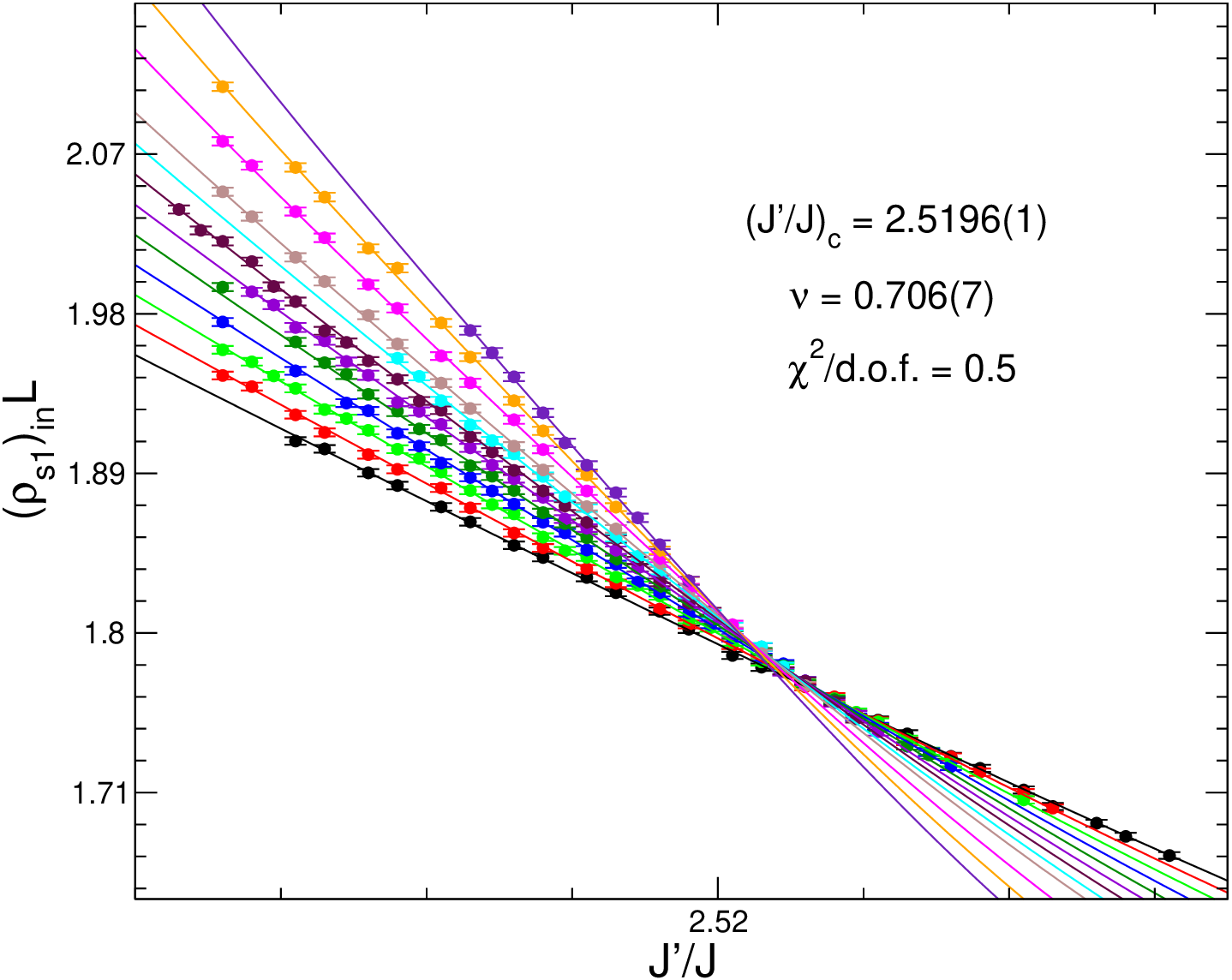}
\end{center}\vskip-0.5cm
\caption{Fit of interpolated $(\rho_{s1})_{{\text{in}}}L$ data
to Eq.~\ref{FSS}. While the circles
are the numerical Monte Carlo data from the simulations, the solid curves 
are obtained by using the results from the fit.
}
\label{fig3}
\end{figure}

\begin{figure}[ht!]
\begin{center}
\includegraphics[width=0.365\textwidth]{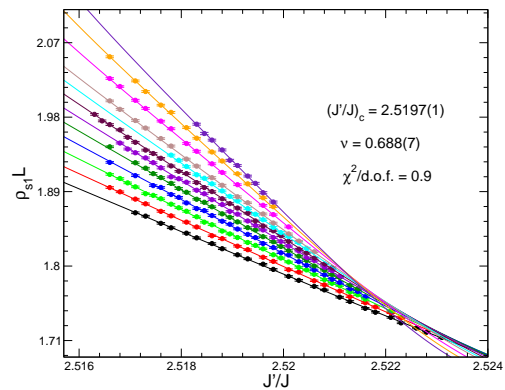}
\end{center}\vskip-0.5cm
\caption{Fit of original $\rho_{s1}L$ data
to Eq.~\ref{FSS}. While the circles
are the numerical Monte Carlo data from the simulations, the solid curves
are obtained by using the results from the fit.
}
\label{fig4}\vskip-0.5cm
\end{figure}

\vskip-0.2cm
To calculate the relevant critical exponent $\nu$ and to determine the
location of the critical point in the parameter space $J'/J$, one useful technique
is to study the finite-size scaling of certain observables. For example,
if the transition is second order, then near the transition, the observable 
$\rho_{si} L$ for $i\in \{1,2\}$ should be described well by the following finite-size scaling 
ansatz
\begin{equation}
\label{FSS}
{\cal O}_{L}(t) = ( 1 + bL^{-\omega} )g_{{\cal O}}(tL^{1/\nu}), 
\end{equation}
where ${\cal O}_{L}$ stands for $\rho_{si}L$, 
$t = (j_c-j)/j_c$ with $j = (J'/J)$, $b$ is some constant,
$\nu$ is the critical exponent corresponding to the correlation length $\xi$ 
and $\omega$ is the confluent correction exponent.  
Finally $g_{{\cal O}}$ appearing above is a 
smooth function of the variable $tL^{1/\nu}$. From Eq.~(\ref{FSS}), one concludes that 
the curves of different $L$ for ${\cal O}_{L}$, as functions of $J'/J$,
should have the tendency to intersect at critical point $(J'/J)_c$ for large $L$.
In the following, we will employ the finite-size scaling formula,
Eq.~(\ref{FSS}), for $\rho_{si} L$ with $i \in \{1,2\}$ to calculate
the critical exponent $\nu$ and the critical point $(J'/J)_c$.
Without losing the generality, in our simulations we have 
fixed $J$ to be $1.0$ and have varied $J'$. Further, the box size used in 
the simulations ranges from $L = 6$ to $L = 64$.
We also use large enough $\beta$ so that the observables studied here 
take their zero-temperature values. Figure \ref{fig0} shows the observables $\rho_{s} L$
and $\rho_{s2} L$ as functions of $J'/J$. The figure clearly indicates the phase 
transition is second order since different $L$ curves
for both $\rho_{s} L$ and $\rho_{s2} L$ tend to intersect at a particular point in 
the parameter space $J'/J$. What is the most striking observation from our results
is that the observable $\rho_{s} L$ receives a much severe correction 
than $\rho_{s2} L$ does. This can be understood from the trend of the crossing among these 
curves of different $L$ in figure \ref{fig0}. Therefore one expects a better 
determination of $\nu$ can be obtained by applying finite-size scaling analysis to $\rho_{s2} L$. 
Before presenting our results,
we would like to point out 
that since data from large volumes might be essential 
in order to determine the critical exponent $\nu$ accurately
as suggested in \cite{Jiang09.2}, we will use the strategy employed
in \cite{Jiang09.2} for our data analysis as well. 
A Taylor expansion of
Eq.~(\ref{FSS}) up to fourth order in $tL^{1/\nu}$ is used to fit the data of $\rho_{s2} L$. 
The critical exponent $\nu$ and critical point $(J'/J)_c$ calculated from the fit using
all available data of $\rho_{s2} L $
are given by $0.6934(13)$ and $2.51962(3)$, respectively. The upper panel of 
figure \ref{fig1} 
demonstrates the result of the fit. Notice both $\nu$ 
and $(J'/J)_c$ we obtain are consistent with the corresponding results found in \cite{Wenzel08}.
By eliminating some data points of small $L$, we can reach a value of 
$0.700(3)$ for $\nu$ 
by fitting $\rho_{s2} L$ with $L \ge 26$ to Eq.~(\ref{FSS}). On the other 
hand, with the 
same range of $L$ ($L \ge 26$), a fit of $\rho_{s} L$ to Eq.~\ref{FSS} leads to $\nu = 0.688(2)$ and $(J'/J)_c = 2.5193(2)$,
both of which are consistent with those obtained in \cite{Wenzel08} as 
well (lower panel in
figure \ref{fig1}). By eliminating more data points of $\rho_{s} L $ with small $L$, 
the values for $\nu$ and $(J'/J)_c$ calculated from the fits are always consistent 
with those quoted above.
What we have shown clearly indicates that one would 
suffer the least correction by considering the finite-size scaling of the observable $\rho_{s2}L$.
As a result, it is likely
one can reach a value for $\nu$ consistent with its $O(3)$ prediction, namely 
$\nu=0.7112(5)$ if
large volume data points for $\rho_{s2}$ are available. Here we do not attempt to carry out
such task of obtaining data for $L > 64$. Instead, we employ the technique of fixing the 
aspect-ratio of spatial winding numbers squared in the simulations. 
Surprisingly, combining the idea of fixing the aspect-ratio
of winding numbers squared and finite-size scaling analysis, even from the observable $\rho_{s1} L$ which
is found to receive the most severe correction among the observables
considered here,
we reach a value for the critical exponent $\nu$ consistent with $\nu=0.7112(5)$ without 
additionally obtaining data points for $L > 64$. The motivation behind 
the idea of fixing the aspect-ratio of spatial winding numbers squared in the simulations
is as follows. 
Intuitively the winding numbers squared $W^2_1$ and $W^2_2$
indicate the ability of the loops moving around $1$- and $2$-directions,
respectively. Further, one can consider the original anisotropic system on the square lattice
as an isotropic system on a rectangular lattice. From these points of view, it is
$\langle W^2_{1}\rangle/\langle W^2_{2}\rangle$, not $(L_2/L_1)^{2}$,
plays the role of the quantity $(L^p_2/L^p_1)^{2}$ for the system,
here $L_{i}$ and $L^p_{i}$ with $i \in \{1,2\}$ are the spatial box size used in the simulations 
and the linear physical length of the system in $i$-direction, respectively.
Indeed it is demonstrated in \cite{Sandvik99} that rectangular lattice is more suitable 
than square lattice for
studying the spatially anisotropic Heisenberg model with different antiferromagnetic couplings 
$J_1$, $J_2$ in $1$- and $2$-directions. The idea of fixing the aspect-ratio of spatial winding numbers squared
quantifies the method used in \cite{Sandvik99}. 
In general for a fixed $L_2$, one
can vary $L_1$ and $J'/J$ in order to reach the criterion of a fixed aspect-ratio of 
spatial winding numbers squared
in the simulations. For our study here,
without obtaining additional data, this method is 
implemented as follows. First of all, we calculate 
$\langle W^2_{1}\rangle/\langle W^2_{2}\rangle$ for the data point at $J'/J = 2.5196$ with $L = 40$
which we denote by $w_f$. Notice since only the aspect-ratio
of the linear physical lengths squared is fixed,
we choose $L^p_1 = L$ for our finite-size scaling analysis.
After obtaining this number,  
a linear interpolation for other data points of $\rho_{s1}$
based on $(w/w_{f})^{(-1/2)}$ is performed in order to reach the
criterion of a fixed aspect-ratio of spatial winding numbers squared in the 
simulations. The $w$ appearing above is the corresponding $\langle W^2_{1}\rangle/\langle W^2_{2}\rangle$ 
for other data points. Further, we keep the number $|w/w_f -1|$ smaller than $0.055$ 
so that the interpolation results are reliable.
A fit of the interpolated $(\rho_{s1})_{\text{in}} L$ data to Eq.~\ref{FSS} with $\omega$ being fixed to its $O(3)$ value ($\omega = 0.78$) leads to
$\nu = 0.706(7)$ and $(J/J)_c = 2.5196(1)$ for $36 \le L \le 64$ (figure \ref{fig3}). 
The subscript " in " appearing above stands for interpolation.
Letting $\omega$ be a fit parameter results in consistent $\nu = 0.707(8)$ and 
$(J'/J)_c = 2.5196(7)$. Further, we always arrive at consistent results with 
$\nu = 0.706(7)$ and $(J'/J)_c = 2.5196(1)$ from the fits using $L > 36$ data. 
The value of $\nu$ we calculate from the fit is in good agreement 
with the expected $O(3)$ value
$\nu=0.7112(5)$. The critical point $(J'/J)_c = 2.5196(1)$ is consistent with that
found in \cite{Wenzel08} as well. To avoid any bias, we perform another analysis 
for the same set of Monte Carlo data without interpolation. By fitting 
this set of original data points to Eq.~\ref{FSS} with a fixed $\omega = 0.78$, 
we arrive at $\nu=0.688(7)$
and $(J'/J)_c = 2.5197(1)$ (figure \ref{fig4}), 
both of which again agree quantitatively with those determined in \cite{Wenzel08}. 
Finally we would like to make a comment 
regarding the choice of $w_f$. In principle one can calculate $w_f$ for any $L$
and for any $J'/J$ close to the critical point. However since it is shown in \cite{Jiang09.2} 
that data points of larger volumes
is essential for a quick convergence of $\nu$, it will be desirable to choose $w_f$ such that
the set of interpolated data contains sufficiently many data points from large volumes. 
For example, using the $w_f$ obtained at $J'/J = 2.5191$ ($J'/J = 2.5196$) with
$L = 40$ ($L = 44$), we reach the results of $\nu=0.704(7)$ and $(J'/J)_c = 2.5196(1)$ 
($\nu=0.705(7)$ and $(J'/J)_c = 2.5196(1)$) from the fit with a fixed
$\omega = 0.78$. These values for $\nu$ and $(J'/J)_c$ agree with what we have obtained earlier.
Interestingly, it seems that the idea of fixing the aspect-ratio of winding numbers squared
determines the critical exponent $\nu$ more accurately than the conventional
method of fixing the aspect-ratio of box sizes in the simulations. 
It would be interesting to 
explore this new method systematically including applying it to the 
study of phase transition for other spatially anisotropic Heisenberg models. 

\section{Discussion and Conclusion}
\label{discussion}\vskip-0.2cm
In this letter, we revisit the phase transition of the spin-1/2 Heisenberg 
model with
a spatially staggered anisotropy. We find that the observable $\rho_{s2} L$
suffers a much less severe correction compared to that 
of $\rho_{s1} L$, hence 
is a better quantity for finite-size scaling analysis. Further, by employing 
the method of fixing the aspect-ratio of spatial winding numbers squared in the 
simulations, we arrive at $\nu=0.706(7)$ for the critical exponent 
$\nu$ which is consistent with the most accurate Monte Carlo $O(3)$ result 
$\nu = 0.7112(5)$ by using only up to $L = 64$ data points derived from $\rho_{s1} L$.

\vskip0.5cm

The simulations 
in this study were performed using the ALPS library \cite{Troyer08} on 
personal desktops. This work is supported in part by funds provided 
by the DOE Office of Nuclear Physics under grant DE-FG02-94ER40818.


\begin{thebibliography}{99}
\bibitem{Wenzel08}
S.~Wenzel, L.~Bogacz, and W.~Janke, Phys. Rev. Lett. {\bf 101}, 127202 (2008).

\bibitem{Sandvik97}
A. W. Sandvik, Phys.~Rev.~B {\bf 56}, 11678 (1997).

\bibitem{Sandvik99}
A. W. Sandvik, Phys.~Rev.~Lett. {\bf 83}, 3069 (1999).

\bibitem{Kim00}
Y.~J.~Kim and R.~Birgeneau, Phys. Rev. B {\bf 62}, 6378 (2000).

\bibitem{Wang05}
L. Wang, K. S. D. Beach, and A. W. Sandvik,
Phys. Rev. B {\bf 73}, 014431 (2006).

\bibitem{Jiang08}
F.-J.~Jiang, F.~K\"ampfer, M.~Nyfeler, and W.-J.~Wiese,
Phys. Rev. B {\bf 78}, 214406 (2008).

\bibitem{Alb08}
A.~F.~Albuquerque, M.~Troyer, and J.~Oitmaa, Phys.\ Rev.\ B
{\bf 78}, 132402 (2008).

\bibitem{Wenzel09}
S.~Wenzel and W.~Janke, Phys.\ Rev.\ B {\bf 79}, 014410 (2009).

\bibitem{Wie94}
U.-J.\ Wiese and H.-P.\ Ying, Z.\ Phys.\ B {\bf 93}, 147 (1994). 

\bibitem{Pardini08}
T.~Pardini, R.~R.~P.~Singh, A.~Katanin and O.~P.~Sushkov, Phys. Rev. 
B {\bf 78}, 024439 (2008).

\bibitem{Jiang09.1}
F.-J. Jiang, F. K\"ampfer, and M. Nyfeler,
Phys. Rev. B {\bf 80}, 033104 (2009).

\bibitem{Cam02}
M.~Campostrini, M.~Hasenbusch, A.~Pelissetto, P.~Rossi,
and E.~Vicari, Phys.~Rev.~B {\bf 65}, 144520 (2002).

\bibitem{Jiang09.2}
F.-J.~Jiang and U.~Gerber, J. Stat. Mech. P09016 (2009).

\bibitem{Troyer08}
A.~F.~Albuquerque et.~al, Journal of Magnetism and Magnetic 
Material 310, 1187 (2007).




\end{thebibliography}
\end{document}